# Supercurrent in a room temperature Bose-Einstein magnon condensate

Dmytro A. Bozhko,[1,2] Alexander A. Serga,[1] Peter Clausen,[1] Vitaliy I. Vasyuchka,[1] Frank Heussner,[1]
Gennadii A. Melkov,[3] Anna Pomyalov,[4] Victor S. L'vov,[4] and Burkard Hillebrands[1,*]

[1]*Fachbereich Physik and Landesforschungszentrum OPTIMAS, Technische Universität Kaiserslautern, 67663 Kaiserslautern, Germany*
[2]*Graduate School Materials Science in Mainz, Kaiserslautern 67663, Germany*
[3]*Faculty of Radiophysics, Electronics and Computer Systems, Taras Shevchenko National University of Kyiv, Kyiv 01601, Ukraine*
[4]*Department of Chemical Physics, Weizmann Institute of Science, Rehovot 76100, Israel*

We report evidence for the existence of a supercurrent of magnons in a magnon Bose-Einstein condensate prepared in a room temperature yttrium-iron-garnet magnetic film and subject to a thermal gradient. The magnon condensate is formed in a parametrically populated magnon gas, and its temporal evolution is studied by time-, frequency- and wavevector-resolved Brillouin light scattering spectroscopy. It has been found that local heating in the focal point of a probing laser beam enhances the temporal decrease in the density of the freely evolving magnon condensate after the termination of the pumping pulse, but it does not alter the relaxation dynamics of the gaseous magnon phase. This phenomenon is understood as the appearance of a magnon supercurrent within the condensate due to a temperature- and, consequently, magnetisation-gradient induced phase gradient in the condensate wave function.

The term supercurrent is commonly assigned to a resistance-free charge current in a superconductor or to a viscosity-free particle current in superfluid $^4$He or $^3$He. More generally, this macroscopic quantum phenomenon relates to the collective motion of bosons (including real atoms [1, 2], Cooper pairs of fermions[3, 4] and bosonic quasiparticles [5–10]) coalescing into the same quantum state and, thus, constituting a specific state of matter described by a single, coherent wave function, a Bose-Einstein condensate (BEC) [11–14]. It is remarkable that any type of supercurrent is induced not by an external driving force but by a *phase* gradient in the BEC wave function. Up to now this prominent effect was only observed at rather low temperatures. However, it is known that a BEC can be prepared even at room temperature by a strong increase in the density [15] of weakly interacting quasiparticles such as photons [16], plasmon-exciton polaritons [17] or magnons [18, 19]. Here, we report evidence for the existence of a supercurrent in a magnon BEC at room temperature in a ferrimagnetic insulator. The temporal evolution of the magnon BEC formed in a parametrically populated magnon gas was studied by Brillouin light scattering spectroscopy (BLS) [20, 21]. It has been found that heating in the focus of a probing laser beam locally enhances the decay of the magnon condensate but does not alter the relaxation dynamics of gaseous magnons. Our understanding of this effect is based on the temperature induced spatial variation in the saturation magnetization and, thus, on the variation in the local magnon frequencies across the heated sample area. Because the magnon condensate is coherent across the entire heated area, a spatially varying phase shift is imprinted into its wave function. This spatial phase gradient propels a magnon supercurrent that flows to the BEC areas around the focal spot and decreases, thus, the BEC density in the probing point. The detailed quantitative agreement between our experiments and a rate-equation model, which considers the temperature induced supercurrent jointly with a dispersion-caused magnon counter-current, provides a compelling proof for a supercurrent in a room-temperature Bose-Einstein magnon condensate.

We have studied the appearance of a magnon supercurrent by observing the temporal evolution of a magnon BEC in an in-plane magnetized single crystal yttrium iron garnet (YIG, $Y_3Fe_5O_{12}$) film. YIG is a ferrimagnetic dielectric material with unique low magnon damping [22] allowing both for efficient thermalization of externally injected magnons and for a several hundred nanosecond-long BEC lifetime. In order to achieve Bose-Einstein condensation [19, 23], magnons are in-jected into the YIG spin system via parallel parametric pumping [24], which currently is considered to be the most efficient technique for magnon excitation over a large wavevector range. The process can be described by a splitting of a photon of a pumping electromagnetic wave with nearly zero wavevector and the pumping frequency $\omega_\mathrm{p}$ into two magnons with opposite wavevectors $\pm q$ and frequency $\omega_\mathrm{p}/2$. This magnon injection process is schematically illustrated by the red arrow in Fig. 1, which presents the magnon spectrum calculated for the given experimental conditions.

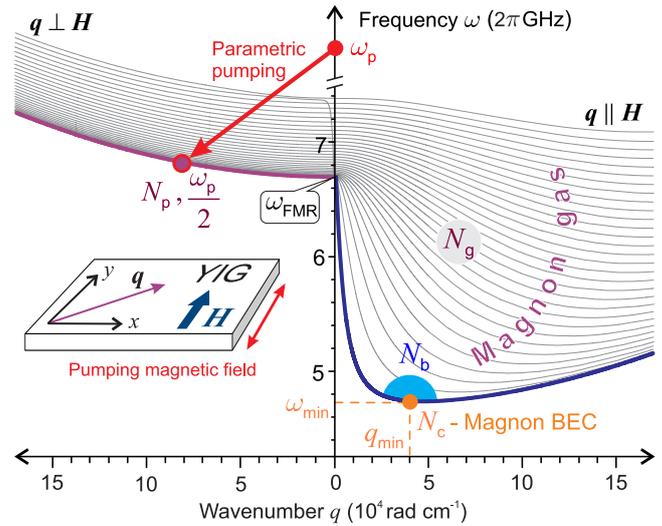

FIG. 1. Magnon spectrum of a 5.6 $\mu$m thick YIG film magnetized in-plane by a bias magnetic field $H = 1690$ Oe shown for the wavevector $q$ perpendicular (left part) and parallel (right part) to the applied field. For both wavevector directions the first 31 thickness modes are shown. The red arrow illustrates the magnon injection process by means of parallel parametric pumping. $N_p$ - total number of parametrically excited magnons at $\omega_p/2$; $N_c$ - number of BEC magnons at $\omega_c = \omega_{min}$; $N_b$ – number of gaseous magnons near $\omega_{min}$ and $q_{min}$; $N_g$ – number of magnons in the parametrically overpopulated gas of magnons below $\omega_p/2$. The inset schematically shows the geometry of the in-plane magnon wavevector $q$, the bias field $H$, and the pumping microwave magnetic field.

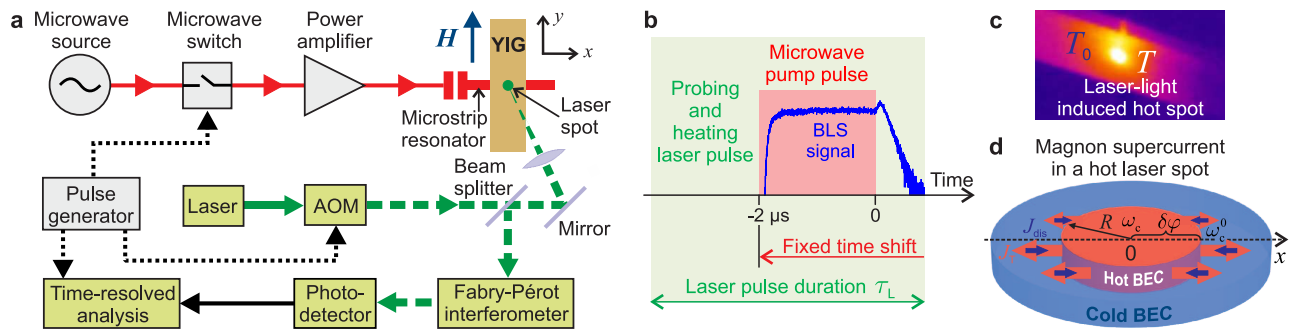

FIG. 2. Experimental setup. **a**, Schematic illustration of the experimental setup. In the upper part of the figure the microwave circuit consisting of a microwave source, a switch and an amplifier is shown. This circuit drives a microstrip resonator, which is placed below the in-plane magnetized YIG film. Light from a solid-state laser ($\lambda = 532$ nm) is chopped by an acousto-optic modulator (AOM) and guided to the YIG film. There it is scattered inelastically from magnons, and the frequency shifted component of the scattered light is selected by the tandem Fabry-Pérot interferometer, detected, and analyzed in time. The time diagram **b** presents the relative timing of the microwave pump pulse, the laser pulse, and the detected BLS signal. **c**, Infrared photo of the YIG sample showing the hot spot in the focus of the probing laser beam. **d**, Schematic illustration of a supercurrent flowing in the magnon BEC subject to a thermal gradient. Red arrows – outflow of the thermally induced supercurrent $J_\mathrm{T}$ from the hot focal spot. Blue arrows – contraflow of the dispersive current $J_\mathrm{dis}$. $\omega_\mathrm{c}$ and $\omega_\mathrm{c}^0$ – magnon BEC phases in the hot and in the cold parts of the sample, respectively. $\delta\varphi$ – temperature induced phase accumulation in the magnon BEC; $R$ – radius of the hot spot.

The strength of the bias magnetic field $H = 1690$ Oe is chosen to allow for magnon injection slightly above the ferromagnetic resonance frequency $\omega_\mathrm{FMR}$, where the parallel pumping process has its highest efficiency [25]. When the injected magnons thermalize through scattering processes conserving both their number and the total energy [26, 27], a BEC may be formed at the lowest energy state of the energy-momentum spectrum even at room temperature [19, 23].

The experimental setup, which consists of a YIG film, a microwave circuit and a BLS system, is schematically shown in Fig. 2a. The YIG sample is 5 mm long and 1 mm wide. The single-crystal YIG film of 5.6 $\mu$m thickness is grown in the (111) crystallographic plane on a gadolinium gallium garnet ($\mathrm{Gd_3Ga_5O_{12}}$) substrate by liquid-phase epitaxy. The microwave pumping circuit comprises a microwave source, a switch, and an amplifier. This circuit drives a 50 $\mu$m wide microstrip resonator, which is placed below the YIG film. The resonator concentrates the applied microwave energy and induces a microwave Oersted field oriented along the bias magnetic field $H$, thus realizing conditions for the parallel parametric pumping mechanism. The pumping pulses of $\tau_\mathrm{P} = 2 \, \mu$s duration are applied with a repetition time of 1 ms allowing for magnetic and temperature equilibration of the YIG film.

In our experiment, a focused laser beam combines the role of the magnon probe in the BLS experiment with the role of the local sample heater. The heating time is adjusted by chopping the probing laser beam using an acousto-optic modulator (AOM). The laser pulse duration $\tau_\mathrm{L}$ is varied between 6 $\mu$s and 80 $\mu$s. By changing $\tau_\mathrm{L}$ the setup allows one to change the heating time interval before the application of the microwave pumping pulse and, consequently, to control the sample temperature in the probing point (see Fig. 2b). The modulated probing beam is focused onto the surface of the YIG film sample in the middle of the microstrip resonator (see Fig. 2a and 2c), where it has a maximum peak power of 9.5 mW. The focal spot has a radius $R \approx 10 \, \mu$m and, thus, is approximately twice smaller than the 50 $\mu$m wide parametric pumping area. The scattered light is deflected by the beam splitter to the tandem Fabry-Pérot interferometer, which selects the frequency shifted component of the scattered light. The intensity of the inelastically scattered light, which is proportional to the magnon density in the probing point, is detected and analyzed in time.

In order to understand the of the magnon BEC, one first needs to separate the effects caused by the spatially uniform change of the sample temperature from those caused by the formation of a temperature gradient. Therefore, in a first experiment, we combined BLS probing at low laser power with an uniform heating of the YIG sample by a hot air flow. The sample temperature was measured by an infrared camera. Magnons were detected near their spectral energy minimum in a frequency band of 100 MHz spectral width.

Figure 3a shows the typical dynamics of the magnon BEC in this case. During the action of the pump pulse the magnon density increases and saturates. After the parametric pumping is switched off, the magnon density jumps up due to the previously reported intensification of the BEC formation in the freely evolving magnon gas [23]. Afterwards the magnon density exponentially decreases due to the conventional spin-lattice relaxation mechanisms. This behavior is common for all temperatures in our experiment. Some decrease in the steady-state magnon density observed at higher temperatures can be related both to higher magnon damping [22] and to lower efficiency of the parametric pumping in this case. The latter effect is straightforward: Due to a lowered saturation magnetization at higher temperatures the parametrically populated spectral magnon branch shifts towards lower frequencies, see Fig. 3b. This shift increases the wavenumber of the injected magnons and, consequently, the threshold power of the parametric generation [25].

Now we focus on the temperature-gradient dependent behavior of the magnon condensate. Figure 4 shows the evolution of the magnon density at the bottom of the spin-wave spectrum for four different heating times, i.e. for four values of the temperature gradient, at a laser power of $P_\mathrm{L} = 9.5$ mW. For comparison, the black curve presents the time-dependent magnon dynamics measured using a much lower laser power of 0.4 mW. Similar to Fig. 3a, the BLS signal rises sharply after the microwave pumping pulse is switched off due to the intensification of the BEC formation process [23]. Afterwards, the magnon density decreases. In the case of low-power laser probing this decrease has an exponential form with a characteristic decay time $\tau \simeq 240$ ns, which corresponds well to the conventional values of a linear magnetic decay in YIG films [22] in the vicinity of $\omega_\mathrm{min}$, i.e. for both the magnon BEC and the gaseous magnons near the bottom of the spectrum.

At higher laser power the observed decrease in the magnon density cannot be described by one single exponential function anymore. Two different regimes of the magnon decay are clearly visible from the slopes of the BLS signals presented in Fig. 4. In the first regime, at high BLS intensities, the decay of the magnon density depends on the heating duration, i.e. on the temperature gradient in the laser focal spot. In the second regime, at low BLS intensities, this decay is temperature independent and its rate coincides with the one observed in the low-laser-power experiment. It is remarkable that for all heating durations the transition between these two decay regimes occurs at approximately the same magnon density

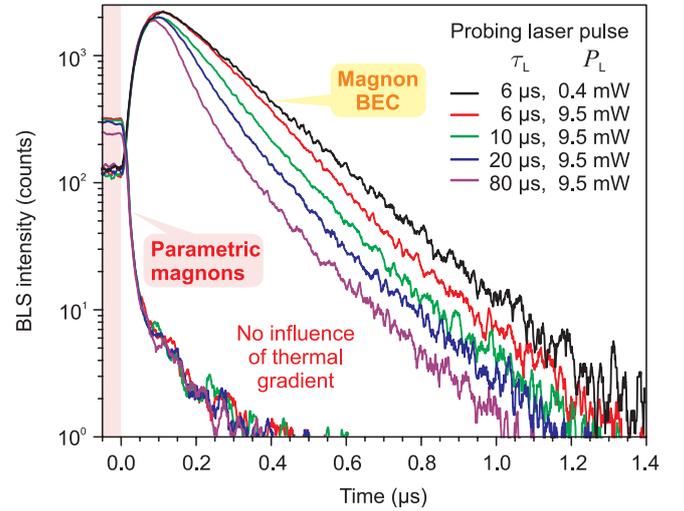

FIG. 4. Temporal dynamics of the magnon BEC under local laser heating. The increase of the laser pulse duration, and thus the heating time, leads to a faster decay of the magnon BEC in the measurement spot, i.e. laser spot. The relaxation time of the residual gaseous magnons at the bottom of the spin-wave spectrum is independent of the heating. The parametrically injected magnons at $\omega_\mathrm{p}/2$ show no temperature related change in their nonlinear relaxation dynamics after the end of the pumping pulse.

level, which can be associated with the transition from the condensed to the gaseous magnon phase.

It is also worth to analyze the evolution of parametrically injected magnons (see Fig. 4), which play the important role of a magnon source for all other spectral states of the magnon gas and, thus, can strongly influence the BEC dynamics. In the beginning, this dense magnon group rapidly decays due to intense four-magnon scattering processes, see e.g. Ref. [28]. This decay closely relates to the jump in the BEC population number shown in Fig. 4. In the course of time, the density of the parametrically excited magnons decreases, four-magnon scattering decelerates and, eventually, the magnon relaxation attains to its intrinsic value corresponding to the decay rate of the gaseous magnons at the bottom of the magnon spectrum. As is clearly seen in Fig. 4, this nonlinear relaxation behaviour is independent of the laser heating and, thus, cannot be the reason for the temperature-gradient-related BEC dynamics.

We now present a model to describe our experimental findings. The magnon BEC is a spontaneously established coherent ground state, which possesses well defined frequency, wavevector, and phase [29]. The local laser heating process locally changes the saturation magnetization and, thus, induces a weak frequency shift $\delta\omega_\mathrm{c}$ between different parts of the magnon condensate. In the course of time this frequency shift results in an increasing phase gradient in the magnon BEC. As a result, a phase-gradient-induced magnon current or, in other words, a magnon supercurrent, flowing out of the hot region of the focal spot, is excited. This efflux reduces the density of the magnon BEC in the probing point. After some time, the enhanced decrease in the magnon density results in the disap-pearance of the condensate, and thus in the disappearance of the supercurrent. Consequently, this leads to the recovery of the conventional relaxation dynamics associated with a residual incoherent gaseous magnon phase.

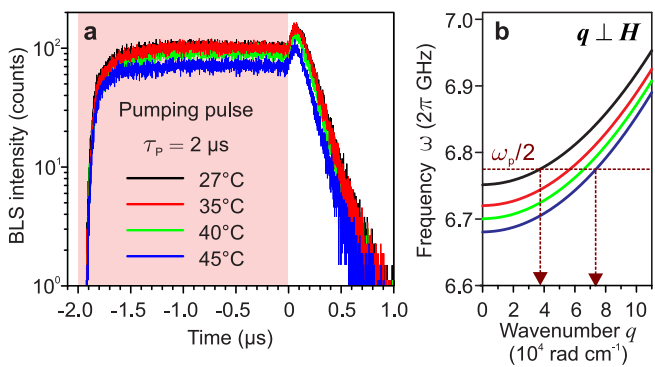

FIG. 3. Temperature-dependent temporal dynamics of the magnon BEC. **a**, Time-resolved BLS intensity for different uniform temperatures of the YIG film. **b**, Influence of the YIG film temperature on the spin-wave dispersion (same color code for temperature as in **a**).

We would like to emphasise that the laser heating process decreases locally the saturation magnetization via thermal excitation of high-energy magnons with terahertz frequencies but practically does not increase the local population of low-energy magnons. As a result, the number of thermal magnons in the low-energy spectral region remains negligibly small in comparison to the number of magnons originating from the parametric pumping process and cannot visibly affect the BEC dynamics. Moreover, the observed temperature-independent relaxation dynamics of the incoherent gaseous magnon phase (see Fig. 4) provides unambiguous evidence for the fact, that the interaction between high- and low-energy magnons is negligibly small: within experimental resolution a diffusively expanding cloud of thermally excited high-energy magnons does not drag the low-energy gaseous magnons out of the heated spot and, therefore, does not contribute to the enhanced local decay of the magnon BEC.

To quantify our interpretation, that a magnon supercurrent is acting, recall that a BEC of magnons can be described by the Gross-Pitaevskii equation for the complex wave function $C(\mathbf{r}, t) = \sqrt{N_c} \exp(i\varphi)$, where $N_c(\mathbf{r}, t)$ and $\varphi(\mathbf{r}, t)$ are the magnon number density per unit volume and the BEC phase, respectively. This equation is the well known equation for the envelope of a narrow wave packet (see e.g. Refs. [28, 30]). It conserves the total number of condensed magnons $\mathcal{N}_c = \int N_c(\mathbf{r}, t)\, d\mathbf{r}$ and leads to the continuity equation for the BEC density $N_c(\mathbf{r}, t)$:

$$\frac{\partial N_c}{\partial t} + \nabla \cdot \mathbf{J} = 0, \quad J_i = \frac{N_c}{2} \sum_j \frac{\partial^2 \omega(\mathbf{q})}{\partial q_i \partial q_j} \frac{\partial \varphi}{\partial r_j}. \quad (1)$$

Here $\mathbf{J}(\mathbf{r}, t)$ is a supercurrent formed by the BEC magnons. In the $(x, y)$-plane of a YIG film (see Fig. 2d) it has two components

$$J_x = N_c D_x \partial\varphi/\partial x, \quad J_y = N_c D_y \partial\varphi/\partial y, \quad (2)$$

where $D_x = d^2\omega(\mathbf{q})/(2\, dq_x^2)$ and $D_y = d^2\omega(\mathbf{q})/(2\, dq_y^2)$ are the anisotropic dispersion coefficients. At our experimental conditions $D_x \simeq 21 D_y$ and, thus, $J_x \gg J_y$. This allows us to neglect $J_y$ in the global balance of the magnon numbers and to simplify the problem to a 1D case along the $x$-axis, see Fig. 2d.

There are two reasons for the $x$-dependence of the BEC phase $\varphi$ in our experiment. The first one is the already mentioned temperature dependence of $\omega_c$. Within the hot spot of radius $R$ centered at $x = 0$ (i.e. for $|x| < R$) the temperature $T(x)$ is higher than the temperature $T_0$ of the rest of the film. (see Fig. 2d). Since in an in-plane magnetized YIG film $d\omega_c(T)/dT < 0$, the BEC frequency in the spot is smaller than outside: $\delta\omega_c(x) = \omega_c(T(x)) - \omega_c(T_0) < 0$. Correspondingly, the phase accumulation $\delta\varphi(x) = \delta\omega_c(x)t$ inside of the spot is smaller than in the surrounding cold film. Therefore, the phase gradient $\partial\delta\varphi(x)/\partial x$ is positive for $x > 0$ and negative for $x < 0$. It means that a thermally induced supercurrent flows out from the spot (mostly in $x$-direction) as it is shown by red arrows in Fig. 2d:

$$J_T = \text{sign}(x) N_c D_x \frac{\partial(\delta\omega_c t)}{\partial x}. \quad (3)$$

This outflow decreases the magnon BEC density $N_c(x)$ in the spot, $|x| < R$, with respect of that in the cold film, where $N_c(x \gg R) = N_c^0$.

Spatial deviations in the density $N_c(x)$ of the magnon condensate constitute the second reason for the variation of its phase $\partial\varphi/\partial x \neq 0$. It results in an additional contribution to the supercurrent given by Eq. (1), which can be named a "dispersive" current $J_{\text{dis}}$. In order to estimate $J_{\text{dis}}$ notice that the 1D Gross-Pitaevskii equation has the self-similar solution $C(x, t) = \mathcal{F}(\xi)$ with $\xi = x^2/(D_x t)$, where the function $\mathcal{F}(\xi)$ satisfies the ordinary differential equation $\xi d\mathcal{F}/d\xi = i(d\mathcal{F}/d\xi + 2\xi d^2\mathcal{F}/d\xi^2)$. It describes the well known phenomenon of dispersive spreading of a wave packet with the width $\delta x = \sqrt{D_x t}$ permanently increasing in time. It is worth to notice that exactly the same law is satisfied by the diffusion process with the diffusion current $J_{\text{dif}} = D_x(\partial N_c(x)/\partial x)$. Ignoring the difference between the self-similar profiles in the dispersion and the diffusion processes (it is well below the resolution of our experiment), we can estimate the dispersive current $J_{\text{dis}}$ by its diffusion counterpart:

$$J_{\text{dis}}(x) \simeq -\text{sign}(x)\, D_x \frac{\partial N_c}{\partial x}. \quad (4)$$

This current is directed to the lower $N(x)$ region, i.e. towards the hot spot as shown in Fig. 2d by blue arrows.

A consistent description of the evolution of a parametrically driven magnon system towards BEC may be achieved in the framework of the weak wave turbulence theory [31], which was further developed in Ref. [28] to describe the spin-wave turbulence under parametric excitation. The main tool of this theory is a kinetic equation for the occupation numbers $n(\mathbf{q})$ of waves (magnons in our case). However, this ambitious aim can only be solved in the future. Here we restrict ourselves to the analysis of a crucially simplified version of the kinetic equations – by rate equations for the total number of magnons in particularly chosen parts of the $\mathbf{q}$-space shown in Fig. 1. The first part is the BEC with density $N_c$ (orange dot in Fig. 1). The second one is the magnon gas with density $N_b$ at the bottom of the magnon spectrum in close vicinity to the BEC (blue area $\mathbb{R}_b$ surrounding the orange BEC point in Fig. 1). These parts are directly coupled via four-magnon scattering processes ($\mathbb{R}_b \to$ BEC) and are simultaneously detected due to the finite resolution of our frequency- and wavenumber-resolved BLS setup. The third part is given by the magnon gas area $\mathbb{R}_g$ above $\mathbb{R}_b$. In our simplified model the parametrically injected magnons of density $N_p$ (magenta dot in Fig 1) populate first the gas area $\mathbb{R}_g$ ($N_p \to \mathbb{R}_g$) and afterwards move to the bottom part $\mathbb{R}_b$ of the magnon spectrum due to four-magnon scattering processes between these areas ($\mathbb{R}_g \rightleftarrows \mathbb{R}_b$).

In the rate equations for the magnon numbers $N_g$, $N_b$ and $N_c$ we have to account that the kinetic equations for the four-magnon scattering processes conserve the total number of magnons and that the leading term for the magnon flux from the $j$ to the $i$ sub-system is proportional to $N_j^3$ and, thus, can be written as $A_{ij}N_j^3$ with $A_{ij}$ being dimensional (s$^{-1}$) phenomenological constants.

Thus, the rate equations take the following form:

$$\frac{\partial N_g}{\partial t} = -\Gamma_g N_g + \Gamma_g N_p e^{-\Gamma_0 t} - A_{gb}N_g^3 + A_{bg}N_b^3, \quad (5a)$$

$$\frac{\partial N_b}{\partial t} = -\Gamma_b N_b + A_{gb}N_g^3 - A_{bg}N_b^3$$
$$- A_{bc}(N_b^3 - N_{cr}^3)\Theta(N_b - N_{cr}), \quad (5b)$$

$$\frac{\partial N_c}{\partial t} = -\Gamma_c N_c + A_{bc}(N_b^3 - N_{cr}^3)\Theta(N_b - N_{cr}) - \frac{\partial J}{\partial x}. \quad (5c)$$

Here $\Gamma_g$, $\Gamma_b$ and $\Gamma_c$ are relaxation frequencies of corresponding magnons in the processes that do not conserve magnon numbers (mainly caused by spin-orbit interaction). Their experimental values are close enough and for simplicity are replaced by the same mean value $\Gamma = 6.1$ MHz.

The second term in the right-hand-side of Eq. (5a) represents the external magnon source and is proportional to the number of parametrically pumped magnons $N_p$. The factor $\exp(-\Gamma_0 t)$ models the processes of magnon thermalization after the end of the pump pulse. The terms $\propto A_{ij}$ in Eqs. (5) describe the magnon fluxes between the chosen spectral areas, leading finally to the population of the BEC state. The flux $\mathbb{R}_b \to \mathbb{R}_c$ contains the Heaviside function $\Theta(N_b - N_{cr})$ that involves $N_{cr}$—a critical number of magnons at which the chemical potential $\mu$ of the magnon gas reaches $\omega_{min}$. The function $\Theta(N_b - N_{cr}) = 1$ for $N_b - N_{cr} > 0$ and 0 otherwise. Therefore, for $N_b \leq N_{cr}$ there is no flux of magnons to the condensate. For $N_b > N_{cr}$ this flux appears and an excess of magnons $N_b - N_{cr}$ populates the condensate.

As we have discussed above, the supercurrent $J$ in Eq. (5c) consists of two parts: $J = J_T + J_{dis}$. Their modelling is, probably, the most delicate issue. We adopt the simplest version, assuming a constant temperature gradient from the center ($x = 0$) of the heated spot to $R$:

$$J = -B_s N_c \delta\omega_c t + B_{dis}(N_c^0 - N_c). \quad (6)$$

Here, for the coefficients we have $B_s \sim B_{dis} \sim -D_x/R$.

The results of the numerical solution of Eqs. (5) and (6) for different $T$ in the center of the hot spot and, accordingly, for different $\delta\omega_c(T)$, are shown in Fig. 5 by color solid lines together with the respective experimental data. One clearly sees the good agreement between the theoretical and experimental curves. The initial rise of the theoretical curves occurs faster than in the experiment. This is likely due to our simplified model with only three zones in the $q$-space. We ignore here the multistage character of the $N_p \to \mathbb{R}_g$ magnon transfer.

In a numerical solution of our model we have choosen the initial BEC value $N_c = 0$ because the gaseous magnons do not populate the lowest energy state in the case of powerful pumping [23]. In such cases, no magnon condensation occurs during the pump action and the BEC in the global energy minimum ($\omega_{min}$, $k_{min}$) is formed *after* the termination of the pump pulse. The initial value of $N_b = 135$ (in units of BLS counts) was taken from our experiment (measured value $N_b + N_c$ at $t = 0$ in Fig. 5). The initial values $N_p = 250$ and $N_g = 1000$ are found from the best fit to the reference case of no heating, shown by a black line in Fig. 5.

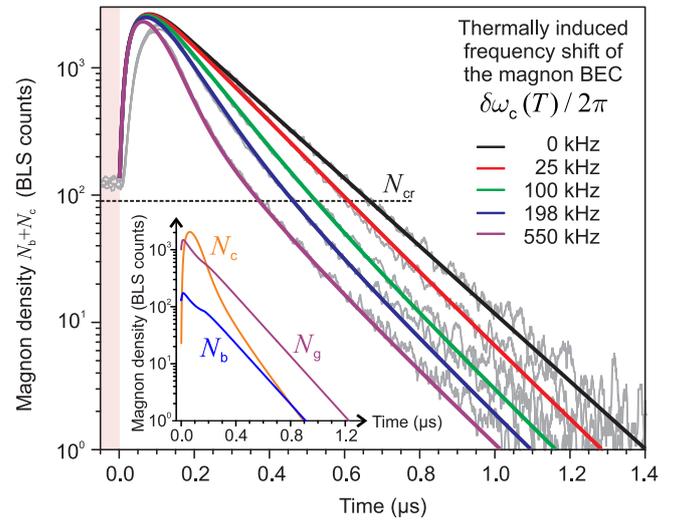

FIG. 5. Theoretically calculated magnon dynamics in a thermal gradient. Theoretical dependencies (colour lines) of the observable magnon densities ($N_c + N_b$) calculated according to Eqs. (5) for different temperature dependent shifts $\delta\omega_c(T)$ of the BEC frequency $\omega_c$ are shown in comparison with the corresponding experimental data (gray curves). The inset shows the calculated temporal behaviour of $N_c$, $N_b$, and $N_g$ for the case of the strongest heating of the magnon BEC.

In the comparison of our model with experiments performed at different spot temperatures $T$, we used only one $T$-dependent parameter $\delta\omega_c(T)$, related to the temperature dependence of the saturation magnetization $M(T)$ of YIG [22]: at standard ambient conditions $dM(T)/dT \simeq -4$ G/K giving $d\omega_c/dT = -116$ kHz/K. For the largest frequency shift ($\delta\omega_c = 550$ kHz), assumed in our calculations, it requires a rather mild heating of the spot by about 4.7 K. This value agrees well with a numerically found temperature rise of 5.7 K in the center of the focal spot. This value was obtained by solving a 3D heat conduction model in the COMSOL Multiphysics simulation package using known values of the light absorption coefficient, heat capacity, duration of the laser pulse, etc.

To summarize, we used Brillouin light scattering spectroscopy to study the time-dependent evolution of the magnon density in the narrow frequency and wavevector vicinity of the bottom of the magnon spectrum. We have found that local laser heating of the probing spot visibly enhances the temporal decrease in the population of the magnon Bose-Einstein condensate freely evolving after the termination of parametric pumping. At the same time, the relaxation dynamics of both gaseous and parametrically injected magnons was not changed. In order to interpret the experimental results we proposed and analyzed a model based on rate equations for the gaseous and condensed magnons, taking into account both the temperature induced magnon supercurrent and the diffusion-like dispersive magnon counterflow via the boundary of the hot spot. In these equations, the position of the magnon energy minima $\hbar\omega_{\min}(T)$ was assumed as the only temperature dependent parameter. The surprisingly high quantitative agreement between the measured data and the analytically modeled magnon evolution at the bottom of the magnon spectrum serves as strong support in favor of the existence of a temperature-induced supercurrent in the Bose-Einstein magnon condensate. Moreover, the occurrence of the supercurrent directly confirms the phase coherency of the observed magnon condensate and fuels expectations of the realization of phase-controlled magnon transport [32, 33], which can be used in future magnonic devices for low-loss information transfer and processing [34]. Therefore, with this demonstration of a magnon supercurrent in a magnon BEC we have opened the door to studies in the general field of magnonic macroscopic quantum transport phenomena at room temperature as a novel approach in the field of information processing technology.

Financial support from the Deutsche Forschungsgemeinschaft (Grant no. INST 161/544-3 within the SFB/TR 49) and from the State Fund for Fundamental Research of Ukraine (SFFR) is gratefully acknowledged. D.A.B. is supported by a fellowship of the Graduate School Material Sciences in Mainz (MAINZ) through DFG funding of the Excellence Initiative (GSC-266). We are also indebted to Y. Tserkovnyak, S. Eggert, A. N. Slavin, V. S. Tiberkevich, K. Nakata, D. Loss, and V. L. Pokrovsky for fruitful discussions. As well we acknowledge I. I. Syvorotka (Scientific Research Company "Carat", Lviv, Ukraine) for supplying us with the YIG film sample.

A.A.S. and B.H. contributed to the experimental idea, planned and supervised the project; D.A.B., P.C., V.I.V. and A.A.S. carried out the experiments; D.A.B. and P.C. contributed to the experimental set-up; D.A.B. carried out the numerical analysis. G.A.M., A.P. and V.S.L. developed the theoretical model, F.H. performed temperature simulations; All authors analysed the experimental data and discussed the results.

---